\pdfoutput=1

\documentclass[8.5pt,twoside,twocolumn]{article}
\usepackage[super,sort&compress,comma]{natbib} 
\usepackage{mhchem}
\usepackage{times,mathptmx}
\usepackage{sectsty}
\usepackage{balance} 

\usepackage{graphicx} 
\usepackage{lastpage}
\usepackage[format=plain,justification=raggedright,singlelinecheck=false,font=small,labelfont=bf,labelsep=space]{caption} 
\usepackage{fancyhdr}
\usepackage{amssymb}   
\usepackage{color} 		

\begin{document}
\thispagestyle{plain}
\fancypagestyle{plain}{
\renewcommand{\headrulewidth}{1pt}}
\renewcommand{\thefootnote}{\fnsymbol{footnote}}
\renewcommand\footnoterule{\vspace*{1pt}%
\hrule width 3.4in height 0.4pt \vspace*{5pt}} 
\setcounter{secnumdepth}{5}

\makeatletter 
\def\subsubsection{\@startsection{subsubsection}{3}{10pt}{-1.25ex plus -1ex minus -.1ex}{0ex plus 0ex}{\normalsize\bf}} 
\def\paragraph{\@startsection{paragraph}{4}{10pt}{-1.25ex plus -1ex minus -.1ex}{0ex plus 0ex}{\normalsize\textit}} 
\renewcommand\@biblabel[1]{#1}            
\renewcommand\@makefntext[1]%
{\noindent\makebox[0pt][r]{\@thefnmark\,}#1}
\makeatother 
\renewcommand{\figurename}{\small{Fig.}~}
\sectionfont{\large}
\subsectionfont{\normalsize} 

\fancyfoot{}
\fancyfoot[RO]{\footnotesize{\sffamily{1--\pageref{LastPage} ~\textbar  \hspace{2pt}\thepage}}}
\fancyfoot[LE]{\footnotesize{\sffamily{\thepage~\textbar\hspace{3.45cm} 1--\pageref{LastPage}}}}
\fancyhead{}
\renewcommand{\headrulewidth}{1pt} 
\renewcommand{\footrulewidth}{1pt}
\setlength{\arrayrulewidth}{1pt}
\setlength{\columnsep}{6.5mm}
\setlength\bibsep{1pt}

\twocolumn[
  \begin{@twocolumnfalse}
\noindent\LARGE{\textbf{Direct measurement of thermophoretic forces}}
\vspace{0.6cm}

\noindent\large{\textbf{Laurent Helden,$^{\ast}$\textit{$^{a}$} Ralf Eichhorn,\textit{$^{b}$} and
Clemens Bechinger\textit{~$^{a,c}$}}}\vspace{0.5cm}


\vspace{0.6cm}

\noindent \normalsize{We study the thermophoretic motion of a micron sized single colloidal particle in front of a flat wall by evanescent light scattering.
To quantify thermophoretic effects we analyse the nonequilibrium steady state (NESS) of the particle in a constant temperature gradient perpendicular to the confining walls.
We propose to determine thermophoretic forces from a ``generalized potential`` associated with the probability distribution of the particle position in the NESS. Experimentally we demonstrate, how this spatial probability distribution is measured and how thermophoretic forces can be extracted with $10~fN$ resolution. By varying temperature gradient and ambient temperature, the temperature dependence of Soret coefficient $S_T(T)$ is determined for $r=2.5~\mathrm{\mu m}$ polystyrene and $r=1.35~\mathrm{\mu m}$ melamine particles. The functional form of $S_T(T)$ is in good agreement with findings for smaller colloids. In addition, we measure and discuss hydrodynamic effects in the confined geometry. The theoretical and experimental technique proposed here extends thermophoresis measurements to so far inaccessible particle sizes and particle solvent combinations.
}
\vspace{2.0cm}
 \end{@twocolumnfalse}
  ]

\section{Introduction}

\footnotetext{\textit{$^{a}$~2. Physikalisches Institut, Universit\"at Stuttgart, Pfaffenwaldring 57, 70550 Stuttgart, Germany.}}
\footnotetext{\textit{$^{b}$~NORDITA, Royal Institute of Technology and Stockholm University, Roslagstullsbacken 23, SE-106 91 Stockholm, Sweden.}}
\footnotetext{\textit{$^{c}$~Max Planck Institut f\"ur Intelligente Systeme, Heisenbergstra{\ss}�e 3, 70569 Stuttgart, Germany.}}

When colloidal particles dispersed in a liquid are exposed to a temperature gradient, they are subjected to thermophoretic forces which drives them towards one side of the gradient. Which side is favored depends on the ambient temperature and the details of particle solvent interactions \cite{Iacopini2006,Wuerger2010}. Thermophoresis has been employed for instance in thermal field flow fractioning to separate colloidal particles \cite{Giddings1993} or in microscale thermophoresis to study protein interactions \cite{Wienken2010}. Furthermore, in hydrothermal pore model systems, a combination of thermophoresis and convection led to an extreme accumulation of nucleotides, RNA and DNA and it is likely that this mechanism played a key role in the evolutionary building up of more complex structures \cite{Baaske2007}.

To quantify thermophoresis of colloids, usually the stationary distribution of a particle ensemble governed by the interplay between thermophoresis in a temperature gradient and Brownian diffusion in a concentration gradient is analyzed by different optical techniques reviewed \textit{e.g.} in \cite{Piazza2008}. This way the Soret coefficients of polystyrene particles of up to $r=1~\mathrm{\mu m}$ radius have been characterized \cite{Duhr2006,Duhr2006b,Braibanti2008}. For larger particles unfeasible long equilibration times and sedimentation effects restrain these ensemble based methods. Even for polystyrene, where sedimentation  can be minimized by matching the density of particles and solvent, micron sized particles are notoriously difficult to measure. 

Here we propose a new strategy to characterize thermophoresis of larger $(r>1~\mathrm{\mu m})$ particles based on a single particle trajectory measurement. It is to a large extend independent of particle buoyancy, so that microparticles composed of so far inaccessible higher density materials like PMMA or melamine can be characterized.

\section{Experimental section}
As experimental method we use total internal reflection microscopy (TIRM, see \cite{Prieve1990,Walz1997} for a review). It is a single particle evanescent light scattering technique capable of measuring the trajectory of a spherical colloidal particle performing its Brownian motion in vicinity of a flat wall. In equilibrium TIRM is well established as a sensitive technique to measure double layer interactions, van der Waals forces and other particle wall interactions. The new idea to characterize thermophoresis is to apply TIRM in a non-equilibrium steady state (NESS) given by a constant temperature gradient and to develop a suitable scheme for data analysis in a NESS.

\begin{figure}[ht]
\centering
  \includegraphics[width=8.3cm]{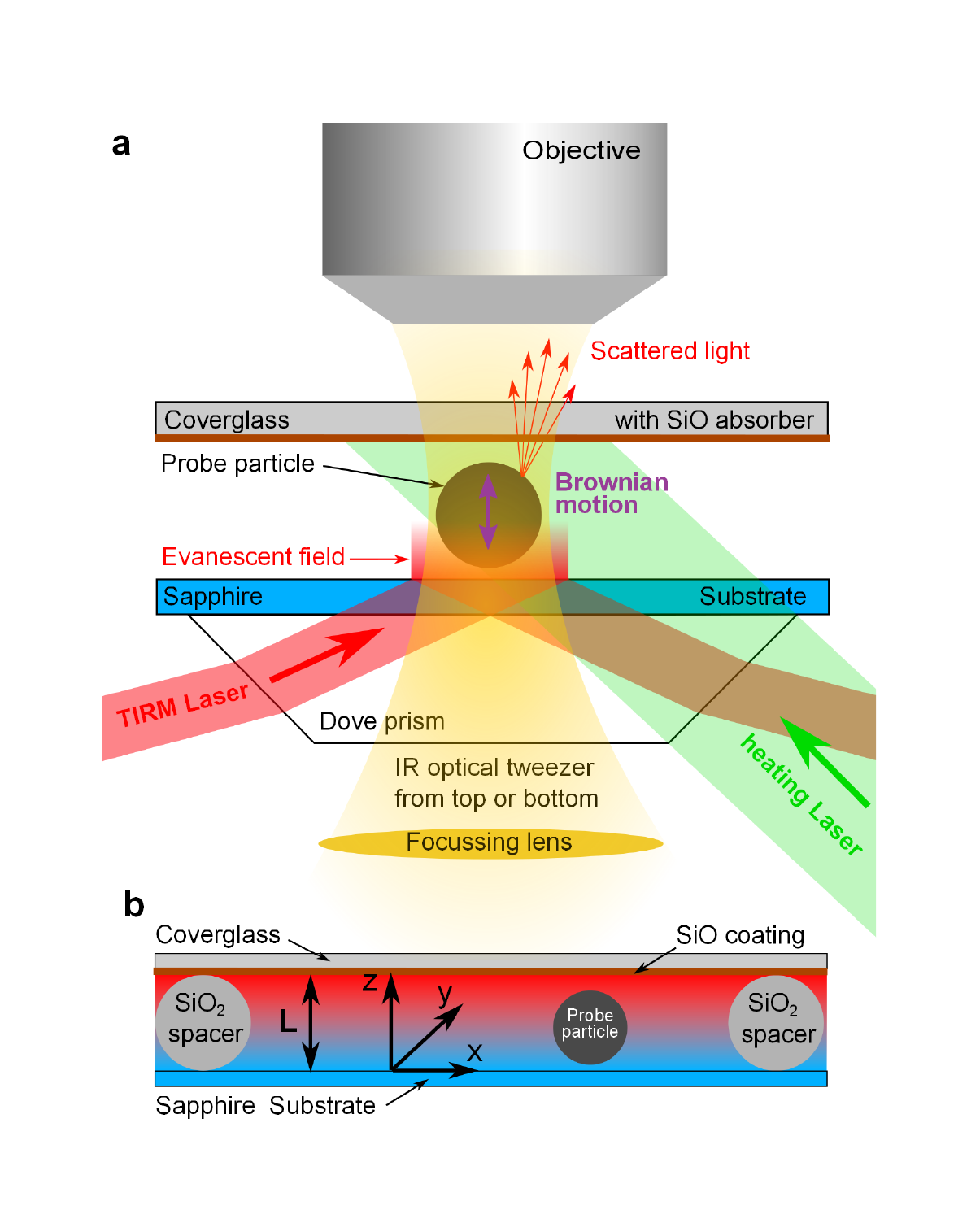}
  \caption{Sketch of experimental set up: a) TIRM principle: The probe particle scatters light from an evanescent field (red) while IR-optical tweezers (yellow) laterally hold the particle in place. In addition the upper coverglas can be heated by a third laser (green) b) Detail of sample cell with spacer particles and temperature gradient. Increasing temperature from blue to red.}
  \label{fig:setup}
\end{figure}

As sketched in Fig.~\ref{fig:setup}a, an evanescent field decaying into the solvent is created by total internal reflection of a laser beam ($\lambda=658~\mathrm{nm}$) at a substrate solvent interface. A single colloidal particle near the substrate will scatter light from this evanescent field. The scattered light intensity is then monitored through microscope optics ($50x, \mathrm{NA}=0.55$ Objective) by a photomultiplier typically for time intervals of 15 to 30 minutes and sampling rates of 1 kHz. 
By converting scattering intensities to particle wall distances the trajectory of the particles motion perpendicular to the substrate is reconstructed. For analysis, data are processed into a histogram that is proportional to the probability density of finding the particle at a specific distance from the substrate. In the following section \ref{sec:theory} we will discuss how to interpret this probability density in a NESS. 
To restrict the lateral diffusion of probe particles and to exert additional light pressure onto the particle, two IR optical tweezers acting either from top ($\lambda=1064~\mathrm{nm}$) or bottom ($\lambda=1070~\mathrm{nm}$) onto the particle are implemented. Both tweezers are only slightly focused such that no detectable gradient forces act perpendicular to the surface. 

For thermophoresis measurements, a stable temperature gradient has to be established in the sample cell. This is accomplished by a thin cell design (Fig.~\ref{fig:setup}b) that allows considerable temperature gradients at only moderate temperature changes. It consists of a $1~\mathrm{mm}$ thick sapphire substrate to ensure good thermal coupling to a heat sink, colloidal silica particles of $7.7~\mathrm{\mu m}$ diameter as spacers to define the cell hight $L$, and a $1\mathrm{mm}$ thick cover glass coated with about $1.8~\mathrm{\mu m} ~ SiO$ layer on the inner side of the cell. The thermally evaporated $SiO$ coating serves as an ($\approx80\%$) optical absorber for green laser light ($\lambda=532~\mathrm{nm}$) while the TIRM detection wavelength ($\lambda=658\mathrm{nm}$) and larger wavelength of optical tweezers are $>95\%$ transmitted. Thus the upper part of the cell above the probe particle can be heated by a slightly focused laser beam ($\approx 200~\mathrm{\mu m}$ beam waist) and the temperature at the absorber is tunable by laser power.

In each sample cell the temperature increase due to laser heating is estimated by the onset of water evaporation \textit{i.e.} bubble formation at high laser powers ($\approx 1~\mathrm{W}$). During this procedure care was taken not to superheat the water by using probe or spacer particles as nucleation sites. The procedure was crosschecked by determining the demixing temperature ($34.0 ^{\circ}\mathrm{C}$) of a critical water 2,6-lutidine mixture in an identically build cell, where due to the spinodal demixing process superheating is impossible\cite{Grattoni1993,Hertlein2008}. Good agreement between both temperature estimates was found. Depending on the actual cell in use, heating of $0.10$ to $0.15~\mathrm{K/mW}$ at the upper cover glass is typical for the experiments. 

For further analysis, heat distribution within the cell was modelled and computed by finite elements methods (COMSOL multiphysics using heat transfer model) taking into account the heat conductance and capacity of prism, substrate, water layer, particle, coverglass and $1~\mathrm{mm}$ air above the cell. The copper housing of the cell was modelled as constant temperature boundary condition while for laser heating a gaussian heating power profile within the SiO layer was assumed.
The numerical results show that the temperature gradient is constant for all particle substrate distances $z$ sampled in the experiments and confirm a linear dependence of the temperature gradient on the laser power.
However, since heat conduction of the sapphire substrate is not infinite, the temperature of the substrate at the sapphire-water interface $T_s$ is not independent of laser heating and the temperature gradient. The change of $T_s$ is, depending on details of cell design, about $1/3$ of the total temperature rise and has to be considered in data analysis.

Due to the thermal expansion of water, the upper part of the water layer in the cell is less dense than the lower part. This together with the thin cell design effectively suppresses possible thermal convection within the cell. We have carefully checked that for the range of heating powers applied during the measurements thermal convection did not occur. Extending the above mentioned numerical simulations to include thermal expansion and the flow within the water layer (COMSOL nonisothermal flow model), we confirm that fluid velocities are below $0.1~\mathrm{nm/s}$ 
at maximum applied temperature gradient. Corresponding stokes drag on the particle is below $0.01~\mathrm{fN}$ and thus orders of magnitude below TIRM force resolution. Signs of convection were actually only observed for extremely high laser powers leading to bubble creation.

As probe particles we used $r=2.5~\mathrm{\mu m}$ radius polystyrene (Type 4205A, Duke Standards, Thermo Scientific, USA) and $r=1.35~\mathrm{\mu m}$ COOH-functionalized melamine (MF-COOH-S1285, Microparticles GmbH, Germany) particles. Prior to use, dispersions were repeatedly washed with Millipore water and otherwise used as supplied. During assembly the cell is cleaned in plasma cleaner and filled with a highly diluted aqueous dispersion of probe particles. The edges of the cell are sealed with UV curable glue. The whole cell is matched to a glass prism with immersion oil and housed in a copper frame with windows above the cell and below the prism to allow optical access to the sample. The copper frame also contains water pipes connected to a flow thermostat such that the cells ambient temperature can be adjusted in the range of $5\ldots 55^{\circ}\mathrm{C}$.

\section{Model and theory} \label{sec:theory}
We describe the probability density $p=p(x,y,z,t)$ for finding a particle
at position $(x,y,z)$ 
in the sample cell at time $t$ by the Fokker-Planck equation
$\frac{\partial p}{\partial t} + \nabla J = 0$,
with the probability current $J=(J_x,J_y,J_z)$.
In the present experimental setup,
$J$ contains contributions
from deterministic forces (particle-wall interactions and
external forcing by optical tweezers and gravity),
thermophoretic drift and thermal noise effects.
We define the coordinate system in such a way, that the $x$-$y$ plane coincides with the inner boundary of
the sapphire substrate and the $z$-axis is oriented along the temperature
gradient (see Fig.~\ref{fig:setup}b); the two walls (sapphire substrate and cover glass)
with their different temperatures are thus located at $z=0$ and $z=L$. Since the optical tweezers confine the particle motion in the $x$ and $y$ directions, and we are only interested in the motion perpendicular to the wall ($z$ direction), we can integrate out the $x$ and $y$ components in the Fokker-Planck equation to obtain
\begin{equation}
\frac{\partial p}{\partial t} + \frac{\partial J_z}{\partial z} = 0 \, ,
\end{equation}
where now $p=p(z,t)$ and $\partial J_z/\partial z|_{z=0} = \partial J_z/\partial z|_{z=L} = 0$
(reflecting boundary conditions).

The probability current $J_z$ along the temperature gradient can be
written as
\begin{equation}
\label{eq:Jz}
J_z = \frac{f}{\gamma}\,p - p D_T \frac{\partial T}{\partial z} - D \frac{\partial p}{\partial z} \, .
\end{equation}
The force $f$ in the first term contains all particle wall interactions, light forces from the optical tweezers and gravity; $\gamma$ is the viscous friction coefficient.
The second term is the standard phenomenological ansatz for the thermophoretic
drift being proportional to the temperature gradient with 
$D_T$ as a phenomenological coefficient quantifying
the thermophoretic effects \cite{Parola2004,Piazza2008,Wuerger2010}. 
The last term represents thermal diffusion with
Einstein's diffusion constant $D=k_{\mathrm{B}}T/\gamma$.
The details connected with the specific form of
the thermal noise term with the space-dependent diffusion
coefficient being written in front of the gradient $\partial/\partial z$
are discussed in the Appendix.

In the following, we are interested in the non-equilibrium steady state
(NESS) where thermophoretic effects are balanced by external forces and
diffusion currents so that $J_z=0$. The current-free solution of
Eq.~(\ref{eq:Jz}) reads
\begin{equation}
\label{eq:p}
p_{\mathrm{NESS}}(z) = \frac{1}{\cal N} \exp \left[ \int_0^z \left( \frac{f}{kT} - \frac{D_T}{D} \frac{\partial T}{\partial z'} \right) \mathrm{d}z' \right] \, ,
\end{equation}
where we need to keep in mind that $f$, $T$, $D$ and, in general,
also $D_T$ depend on position $z$.
The normalization constant $\cal N$ is chosen such that $\int_0^L p(z) \mathrm{d}z = 1$.

Following the procedure in \cite{Blickle2007},
we define the generalized
``pseudopotential'' \cite{speck06,Blickle2007}
as the negative logarithm of the steady-state distribution,
\begin{equation}
\label{eq:Phi}
\Phi(z) := -\ln p_{\mathrm{NESS}}(z)
= - \int_0^z \left( \frac{f}{k_{\mathrm{B}}T} - \frac{D_T}{D} \frac{\partial T}{\partial z'} \right) \mathrm{d}z' + \Phi_0 \, .
\end{equation}
The offset $\Phi_0 = \ln {\cal N}$ is due to normalization; it is irrelevant
for the spatial dependence and thus will be omitted in the following.
The generalized potential is composed of two space-dependent contributions
\begin{equation}
\label{eq:Phi2}
\Phi(z) = \Phi_f(z) + \Phi_T(z) \, .
\end{equation}
The first part
\begin{equation}
\label{eq:Phif}
\Phi_f(z) = - \int_0^z \frac{f}{k_{\mathrm{B}}T} \, \mathrm{d}z'
\end{equation}
is related to the conservative forces $f$ on
the particle.
It simplifies considerably for small constant temperature
gradients.
Writing
\begin{equation}
\label{eq:T}
T(z) = T_s + \theta z
\end{equation}
with the substrate temperature $T_s=T(z=0)$ and 
temperature gradient $\theta=\partial T/\partial z$,
and assuming that $|T(z=L)-T_s|/T_s = \theta L/T_s \ll 1$,
the term $1/k_{\mathrm{B}}T$ in Eq.~(\ref{eq:Phif}) is given by
$1/k_{\mathrm{B}}T_s$ to lowest order in $\theta L/T_s$.
For small temperature gradients $\theta L/T_s \ll 1$,
the generalized potential $\Phi_f$
is thus directly determined by the equilibrium potential
$V(z) = - \int_0^z f \, \mathrm{d}z'$,
\begin{equation}
\label{eq:Phif_0}
\Phi_f(z) \approx \frac{V(z)}{k_{\mathrm{B}}T_s} \, .
\end{equation}
This relation is exact in case
of an equilibrium situation, \textit{i.e.}\ where there are
no temperature variations over the sample cell, $\theta=0$.

The second part in Eq.~(\ref{eq:Phi}),
\begin{equation}
\label{eq:Phi_T}
\Phi_T(z) = \int_0^z \frac{D_T}{D} \frac{\partial T}{\partial z'} \, \mathrm{d}z' \, ,
\end{equation}
 can be interpreted as the generalized
potential of a ``thermophoretic force''
\begin{equation}
\label{eq:fT}
f_T := -k_{\mathrm{B}}T \frac{\partial \Phi_T}{\partial z} = -k_{\mathrm{B}}T \frac{D_T}{D} \frac{\partial T}{\partial z} \, .
\end{equation}
A simplification based on a small gradient expansion as we
performed it for $\Phi_f$ is not easily possible due to the
\textit{a priori} unknown temperature dependence of the coefficient
$D_T/D$.

It is obvious, however, that in thermal equilibrium with
$\partial T/\partial z = \theta = 0$ we have $f_T=0$ and thus $\Phi_T=0$
(again, up to an irrelevant constant).
Using Eqs.(\ref{eq:Phi2}) and (\ref{eq:Phif_0}), we find
$\Phi = V(z)/k_{\mathrm{B}}T_s$ and, finally, from Eq.(\ref{eq:Phi})
\begin{equation}
\label{eq:V}
V(z) = -k_{\mathrm{B}}T_s \ln p_{\mathrm{eq}}(z) \, ,
\end{equation}
when substituting the NESS density $p_{\mathrm{NESS}}$
by its equilibrium counterpart $p_{\mathrm{eq}}$
as the stationary distribution reached under thermal equilibrium conditions.
Because the equilibrium density $p_{\mathrm{eq}}$ is given
by the Boltzmann factor $\exp(-V/k_{\mathrm{B}}T_s)$,
this confirms that the general non-equilibrium approach Eq.~(\ref{eq:Phi})
is consistent with equilibrium statistical mechanics \cite{Blickle2007}.

\section{Data evaluation}
As already mentioned in section 2, the probability density $p(z)$
for finding the particle at distance $z$ from the substrate can be
extracted from the TIRM scattering intensities.
Performing such a TIRM measurement without applying a temperature
gradient, allows to deduce the potential $V(z)$ via Eq.~(\ref{eq:V})
to quantify particle-wall interactions and
the external forces due to the optical tweezers and gravity.

In our experiments, the temperature gradient
$\theta$ is below $1\,\mbox{K}/\mu\mbox{m}$ so that $\theta L/T_s \lesssim 0.025$ 
(with $L=7.7\,\mu\mbox{m}$ and $T_s = 300\,\mbox{K}$)
is indeed negligibly small.
As shown above [see Eq.~(\ref{eq:Phif_0})],
such small temperature variations do not interfere with
the potential forces due to $V(z)$.
We can therefore determine the ``thermophoretic potential''
$\Phi_T$ from the stationary distribution
$p_{\mathrm{NESS}}$ in a given (weak) temperature gradient
by using [\textit{cf.}~Eqs.~(\ref{eq:Phi}), (\ref{eq:Phi2}), (\ref{eq:Phif_0})]
\begin{equation}
\label{eq:PhiTfromNESS}
\Phi_T(z) = \Phi(z)-\Phi_f(z) = -\ln p_{\mathrm{NESS}}(z) - \frac{V(z)}{k_{\mathrm{B}}T_s} \, .
\end{equation}

A central quantity for characterizing thermophoretic effects
is the Soret coefficient $S_T:=D_T/D$. According to Eq.~(\ref{eq:fT})
it is related to the ``thermophoretic force'' and the 
generalized ``thermophoretic
potential'' by
\begin{equation}
\label{eq:STfromPhiT}
S_T = -\frac{f_T}{k_{\mathrm{B}}T\,\partial T/\partial z} = \frac{\partial \Phi_T/\partial z}{\partial T/\partial z} \, .
\end{equation}
The Soret coefficient can therefore be measured from
the stationary particle distribution $p_{\mathrm{NESS}}$
observed in a weak temperature gradient $\theta$ (with $\theta L/T_s \ll 1$)
by making use of the relations (\ref{eq:PhiTfromNESS}), (\ref{eq:STfromPhiT}),
after $V(z)$ has been determined in an independent
equilibrium measurement [Eq.~(\ref{eq:V})].
We remark that Parola and Piazza in \cite{Parola2004},
suggested a relation analogous to Eq.~(\ref{eq:STfromPhiT})
for determining the Soret coefficient, however derived directly
from hydrodynamic forces on the particle due to thermophoretic effects
and without taking into account other conservative forces.

\section{Results and discussion}
\begin{figure}[h]
\centering
  \includegraphics[width=8.3cm]{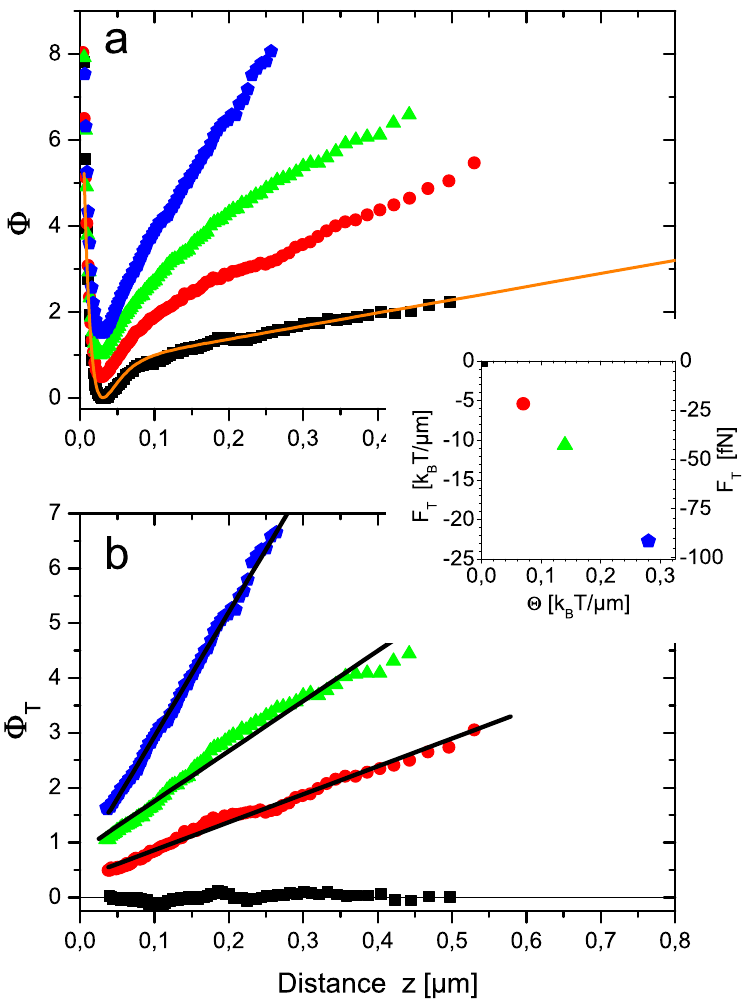}
  \caption{a) Generalized interaction potentials $\Phi(z)$ of a $r=2.5\mu m$ polystyrene probe particle subjected to different temperature gradients $\theta= 0.00, 0.07, 0.14  \mathrm{~and~ }  0.27 K/\mu m$ for black squares, red circles, green triangles and blue pentagons, respectively. Data were taken at $18.7 ^\circ C$ ambient temperature. The orange line $\Phi_{fit}(z)$ is a fit to the equilibrium potential $V(z)/k_BT_s=\Phi_f(z)$ (black squares) as explained in the main text. To improve visibility, potentials have been shifted vertically by $0.5$ with respect to each other. \\
b) ``Thermophoretic potentials'' $\Phi_T(z)$. Same data as in a, but fit to equilibrium potential $\Phi_{fit}(z)$ subtracted. Black lines are linear fits to the data. The inset shows thermophoretic forces, \textit{i.e.}\ negative slopes of the fits as function of temperature gradient $\theta$.
}
\label{fig:forces}
\end{figure} 
Typical TIRM measurements for different temperature gradients are shown in Fig.~\ref{fig:forces}a. Without temperature gradient, the equilibrium potential $V(z)/k_BT_s=\Phi_f(z)$ (black squares) is obtained. For $z<30~\mathrm{nm}$ it exhibits a steep repulsion which is due to screened Coulomb interactions between the negatively charged surfaces of sapphire substrate and sulfate terminated polystyrene particle. 
For larger distances up to about $150~\mathrm{nm}$ a potential well of $0.8~k_\mathrm{B}T$ depth is observed. This can be attributed to attractive van der Waals forces. Towards larger distances the potential has a constant slope that reflects the buoyancy of the particle and additional light forces of optical tweezers. The entire potential is well fitted  by the function $\Phi_{fit}(z)=16.50~\exp(-z/10~\mathrm{nm})-6.53~\exp(-z/21.91~\mathrm{nm})+3.05~mn^{-1}~z+0.35 $ 
displayed as orange line in Fig.~\ref{fig:forces}a.

The first term accounts for electrostatic interactions according to Debye-H\"uckel theory. The fitted Debye length of $10~\mathrm{nm}$ is attributed to the counterions in the thin sample cell. 
The second term describes the van der Waals attraction by an empirical exponential formula given in Eq. 5 of Ref. \cite{Bevan1999}. The last two terms incorporate light pressure and gravity. In the following $\Phi_{fit}(z)$ is used to subtract $\Phi_T(z)$ from $\Phi(z)$ in the presence of
temperature gradients [c.f.~Eq.~(\ref{eq:PhiTfromNESS})]. The result is shown in Fig.~\ref{fig:forces}b for $z \gtrsim 30~\mathrm{nm}$, \textit{i.e.}\ the position of the potential minimum. 
The pure thermophoretic potentials $\Phi_T$ obtained at different temperature gradients are, within errors, linear functions of distance. This means that the probe particle experiences a constant thermophoretic force within the distance range sampled. For bulk measurements this is certainly expected. However, in the vicinity of a surface it is worthwhile to discuss the different contributions to $f_T$ (Eq.~\ref{eq:fT}), in particular due to the hydrodynamic wall effects altering the diffusion coefficient $D(z)$.

\begin{figure}[h]
\centering
  \includegraphics[width=8.3cm]{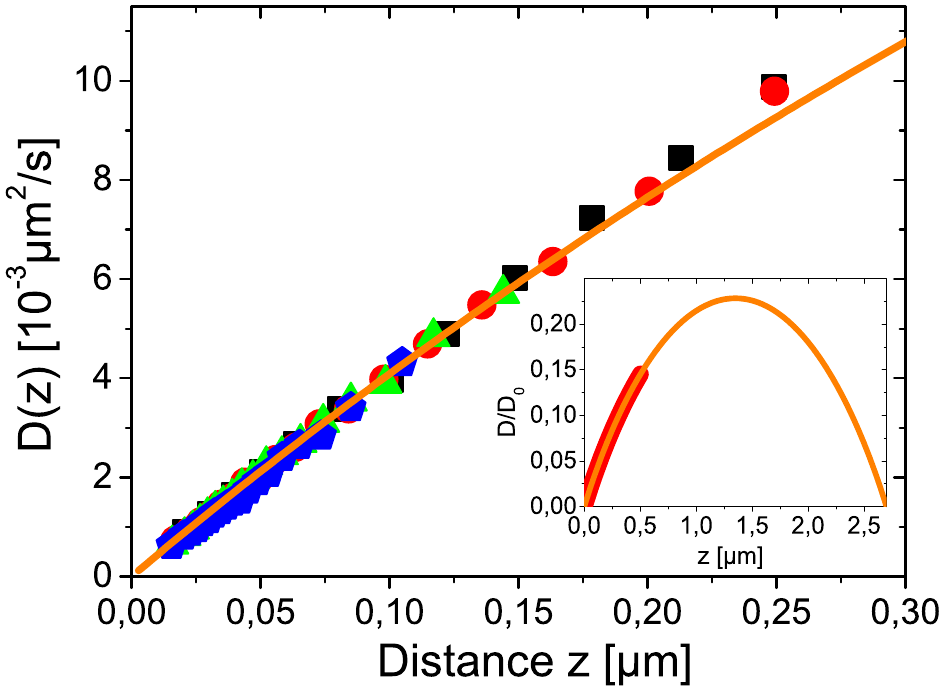}
  \caption{Measured spatial dependence of Brownian diffusion coefficient for different temperature gradients $\theta= 0.00, 0.07, 0.14$ and $0.27\mathrm{K/\mu m}$ for black squares, red circles, green triangles and blue pentagons respectively. $D(z)$ was derived from the same data as used for Fig. \ref{fig:forces} and symbols correspond. The orange line is a fit to theory given in \cite{Lobry1996} calculated for $r=2.5~\mathrm{\mu m}$ particle in a $L=7.7~\mathrm{\mu m}$ slit with $D_0$ as only fitting parameter. The inset shows the theoretical prediction for $D(z)$ on a larger scale with the range of experimental data marked as thick red line.}
\label{fig:diffusion}
\end{figure}

Close to a wall the hydrodynamic friction coefficient $\gamma$ of a spherical colloidal particle is changed drastically (it even becomes anisotropic) compared to the bulk value $\gamma_0=6\pi\nu r$ ($\nu$ being the fluid viscosity) given by Stokes' solution. By Einstein's relation $D(z)=k_{\mathrm{B}}T(z)/\gamma(z)$ also the Brownian diffusion coefficient for diffusion perpendicular to the wall $D(z)$ acqires a pronounced distance dependence.
Exploiting, as before, that the temperature variations over the sample cell are small in our experiments,
$\theta L/T_s \ll 1$, we can neglect the space-dependence of temperature and find in lowest order
$D(z)=k_{\mathrm{B}}T_s/\gamma(z)$.
The effects of hydrodynamic corrections in the friction and (normal) diffusion coefficient
close to the walls can therefore be considered to be unaffected by the (weak) temperature gradients.
The theoretical prediction for the normal diffusion coefficient is displayed in Fig.~\ref{fig:diffusion}
as orange line.
It has been calculated by Brenners formula \cite{Brenner1961} and its extensions for a thin slit between two walls \cite{Lobry1996,Benesch2003}, which represents the actual experimental situation. The inset predicts that in the experimental slit geometry diffusion is at maximum only $23\%$ of the bulk diffusion coefficient
$D_0=k_{\mathrm{B}}T_s/(6\pi\nu r)$.

Experimentally, the spatially resolved diffusion coefficient can be extracted from dynamical analysis of TIRM-data according to a procedure described in \cite{Oetama2005}. It is shown as symbols in Fig.~\ref{fig:diffusion}. For all temperature gradients data are in remarkable agreement. This proves that the Brownian motion is completely independent of thermophoretic effects. Since $\partial \Phi_T/\partial z$ is constant within the experimentally accessible distance range, it follows from Eq. \ref{eq:fT} that $D_T(z)$ has the same distance dependence as $D(z)$ and the Soret coefficient $S_T=D_T(z)/D(z)$ is constant \textit{i.e.} independent of $z$. 

The magnitude of the thermophoretic force, \textit{i.e.}\ the negative slopes of the fitted (black) lines in Fig.~\ref{fig:forces}b, also depends linearly on the temperature gradient, as shown in the inset of Fig.~\ref{fig:forces}b. This experimental observation confirms the usual assumption that thermophoretic velocities (and forces) are proportional to the temperature gradient and $S_T$ is independent of the temperature gradient. 
It also imparts an impression on the magnitude of thermophoretic forces which are here in the range of about $20 - 100~\mathrm{fN}$.

\begin{figure}[h]
\centering
\includegraphics[width=8.3cm]{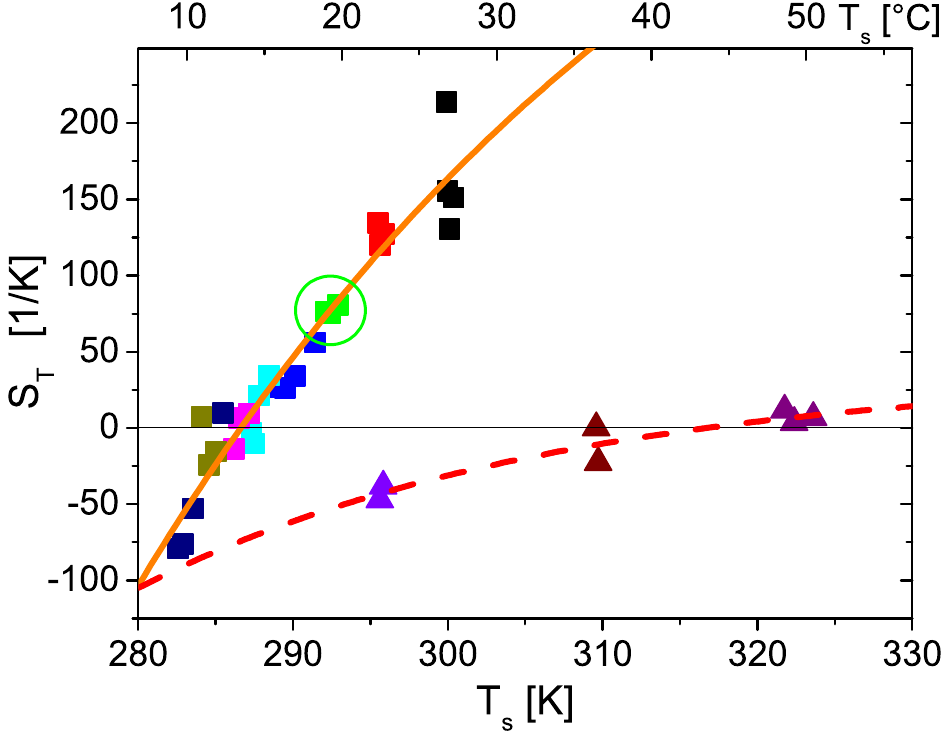}
  \caption{Temperature dependence of Soret coefficient obtained for different ambient temperatures (indicated by different colors), temperature gradients and particle types (indicated by symbol types). Square symbols are for a $r=2.5~\mathrm{\mu m}$ polystyrene probe particle. In particular green squares marked by a circle correspond to the data shown in Fig. \ref{fig:forces} . Triangles are for a $r=1.35~\mathrm{\mu m}$ melamine particle. The spread of the symbols for equal ambient temperature along the temperature axis is due to increasing substrate temperatures $T_s$ for increasing temperature gradients as mentioned in the main text. The lines are fits to Eq. \ref{eq:soret}. For polystyrene particles (full orange line) $S_T^{\infty}=583~\mathrm{K}^{-1}, T^*=286.6~\mathrm{K},$ $T_0=40.6~\mathrm{K}$ and melamine particles (dashed red line) $S_T^{\infty}=36.0~\mathrm{K}^{-1}, T^*=316.7~\mathrm{K}, T_0=27.3~\mathrm{K}$. }
\label{fig:soret}
\end{figure}

While $S_T$ does not depent on temperature gradient, a characteristics of thermophoresis is its pronounced dependence on the absolute temperature. It has been shown that for many substantially different dispersed systems, like polystyrene nanoparticles, Lysozyme micelles, DNA etc. \cite{Braibanti2008, Putnam2007, Iacopini2003, Iacopini2006} the temperature dependence of the Soret coefficient follows a common empirical fitting formula introduced by Iacopini and Piazza \cite{Iacopini2003},
\begin{equation}
S_{T}(T)=S_T ^{\infty} \left[ 1-\exp \left( \frac{T^* -T}{T_0}     \right) \right] 
  \label{eq:soret} 
\end{equation}
with system specific constants $S_T ^{\infty}, T^*$ and $T_0$.

Figure \ref{fig:soret} shows the temperature dependence of $S_T$ for probe particles of different material. 
As an example of particles that due to their size and strong sedimentation (density $\varrho_{MF}=1.51~\mathrm{g/cm}^3$) are difficult (if not impossible) to access with other techniques, we present data for $r=1.35~\mathrm{\mu m}$ melamine particles (triangles) with carboxyl terminated surface. Qualitatively, they follow the temperature dependence given by Eq. (\ref{eq:soret}) with a remarkably high $T^*=43.5~^{\circ}\mathrm{C}$. Hence at room temperature they have a negative $S_T$ and belong to the rare group of thermophilic particles.

$S_T(T)$ for $r=2.5~\mathrm{\mu m}$ polystyrene particles also accords with Eq.~(\ref{eq:soret}). Even the temperature of sign reversal is in good agreement with previous results for considerably smaller polystyrene particles of $53~\mathrm{nm}$ to $253~\mathrm{nm}$ radius where $T^*=286.6\mathrm{K}~\widehat{=}~13.4^{\circ}\mathrm{C}$ is reported \cite{Braibanti2008}. Comparing the absolute values of $S_T$ for the $r=253~\mathrm{nm}$ particles in these measurements and for the 10 times larger polystyrene particles in the present experiment, we find in our experiments that $S_T$ is two orders of magnitude larger. This certainly exceeds the linear prediction for the size dependence of $S_T$ found for smaller particles measured in \cite{Braibanti2008} and expected from theories of thermophoresis based on a flowfield around the particle \cite{Anderson1989,Piazza2008,Wuerger2010}. While in our experiments no salts or surfactants where added, dispersions of Ref. \cite{Braibanti2008} were treated with Triton X100 surfactant and experiments where performed in a density matched $H_2O:D_2O$ mixture containing $1~\mathrm{mM}$ Tris-$HCl$ pH~$7.8$ buffer and $10~\mathrm{mM}$ $NaCl$. These different condition certainly question the comparability of both measurements even though $T^*$ known to depend sensitively on the particle-solvent interface \cite{Iacopini2006} is quite similar.

In experiments with $r=1.25~\mathrm{\mu m}$ polystyrene particles (data not shown) measured under same conditions as the $r=2.5~\mathrm{\mu m}$ particles , we find a 4-5 times smaller $S_T$. Thus our findings rather suggest a quadratic size dependence of $S_T$ found in \cite{Duhr2006,Duhr2006b}. Such a quadratic dependence is theoretically supported by thermodynamic arguments \cite{Duhr2006,Duhr2006b}. However our combination of $S_T$, particle size and temperature gradients $S_T r \theta \approx 5\ldots 100$ largely exceeds the applicability of this theory $S_T r \theta \ll 1$.

Finally, if we presume a driving mechanism for thermophoretic motion that creates a flowfield around the particle\cite{Anderson1989,Piazza2008,Wuerger2010}, it is clear, that the pronounced influence of confining surfaces on hydrodynamics shown in Fig.~\ref{fig:diffusion} will not only influence the Brownian diffusion but also the thermophoretic propulsion itself \cite{Morthomas2010,Yang2013a}. On a length-scale of the order of the particle radius, \textit{i.e.}\ for distances considerably larger than those sampled in Fig.~\ref{fig:forces}, this would lead to a spatial variation of thermophoretic forces. If this is indeed the case, comparison with bulk measurements and furthermore the concept of $S_T$ (which should be a particle property, independent of distance from boundaries) is inappropriate in confined geometries. A distance dependence of thermophoretic forces could also explain measurements showing a size dependence of $S_T$ which is stronger than linear within this hydrodynamic concept. These issues will be in the focus of our further research.

\section{Conclusions}
In conclusion we have demonstrated the applicability of TIRM in NESS-systems and pointed out a suitable scheme for data analysis. The origin of nonequilibrium is a temperature gradient that drives a thermophoretic motion of micron sized colloidal particles. Thermophoretic forces have been defined and directly measured with a precision in the $10~\mathrm{fN}$ range. In good qualitative agreement with existing measurements for smaller polystyrene particles we measured the temperature dependence of $S_T$ for polystyrene particles. 
Influences of the confined geometry on $D$, $D_T$ and $S_T$ were discussed. This might have important consequences for the applicability of thermophoresis in microfluidic devices. Furthermore we demonstrated the potential of the method to characterize micron sized particles with higher densities with respect to their thermophoretic properties by determining $S_T (T)$ for melamine particles in agreement with the empirical formula by Iacopini and Piazza \cite{Iacopini2003}. The new technique not only makes a larger variety of microparticles accessible to thermophoretic measurements, but also opens up a route for generalization to different solvents like alcohols, hydrocarbons and other unpolar liquids. This might allow to study the dependence of thermophoresis on detailed particle solvent interaction in the near future.

\paragraph*{Acknowledgements:}\hspace{2mm} We thank Daniele Vigolo and Alois W\"urger for stimulating discussions and Daniel Maier for COMSOL simulations. R.E. acknowledges financial support by the Swedish Science Council under the grant 621-2013-3956.

\section*{Appendix}
In this Appendix, we discuss the details
associated with formulating
overdamped Brownian motion in an inhomogeneous thermal
environment with
space-dependent diffusion coefficient
in the context of thermophoresis.
For simplicity we restrict ourselves to one
spatial dimension which is denoted by $z$ (as in
the main text). We furthermore omit any deterministic forces,
so that the probability current $J_z$ for the particles
in a dilute suspension consists of a thermophoretic part
proportional to the gradient of temperature $T=T(z)$ and
a diffusive part proportional to the gradient of
the probability density $p=p(z,t)$,
\begin{equation}
\label{eq:Jz_App}
J_z = - p D_T \frac{\partial T}{\partial z} - D \frac{\partial p}{\partial z} \, .
\end{equation}
This description represents
the standard form used in the thermophoretic
literature \cite{Wuerger2010,Parola2004,Piazza2008}, based
on the reciprocal formulation of heat and particle currents
driven by temperature and density gradients \cite{deGroot85}.
The so-called ``thermal diffusion coefficient'' $D_T$
essentially quantifies the effects of the thermal gradient on
a thin layer at the particle-solvent interface,
where interfacial tension gradients parallel to the temperature gradient
drive thermophoretic particle motion
\cite{Wuerger2010,Piazza2008}.
The strength of the diffusion term is given by
the diffusion coefficient $D=k_{\mathrm{B}}T/\gamma$.

Our main concern in this Appendix is the
specific form of this diffusion term.
In thermophoresis, $T$ and thus  $\gamma$ and $D$ are position-dependent.
It is therefore not clear that $-D\partial p/\partial z$ is
the ``correct'' form to describe particle diffusion in
such inhomogeneous thermal environment
or if additional ``spurious'' drift terms would have to
be added in Eq.~(\ref{eq:Jz_App})
\cite{ryter81,Sancho1982,vanKampen1987,Widder1989,Jayannavar1995,Lau2007,hottovy12}.
This problem is \textit{per se} not related
to thermophoretic mechanisms as sketched above,
but has a different origin rooted essentially in the
mathematical description of Brownian motion.
It is connected
to the so-called It\^o-Stratonovich dilemma for the stochastic
Langevin equation associated with Eq.~(\ref{eq:Jz_App}) \cite{vanKampen1981,vanKampen2007,Sancho2011,Yang2013b}.
In fact, as fluid viscosity changes with temperature,
the friction coefficient $\gamma$ depends
on particle position as well. In our experimental setup, where we measure
thermophoresis close to the substrate surface in a slit geometry (see Fig.~1),
$\gamma$ even acquires an additional
effective dependence on the wall distance $z$ by hydrodynamic effects
\cite{BrennerBook}
\footnote{In the current experimental
situation with temperature variations of at most a few degrees Kelvin within the sample cell, the
corresponding changes in viscosity are below 10\% and thus are
negligible in comparison with the hydrodynamic wall effects, see also Fig.~3.}
In other words, variations of $D$ with position $z$ are the result of
separate dependencies of $T$ and $\gamma$ on $z$.
It is well-known \cite{ryter81,Sancho1982,Jayannavar1995,hottovy12,Yang2013b}
that under such conditions, the correct form of the diffusion term is
$-(1/\gamma)\partial (k_{\mathrm{B}}T\,p)/\partial z$, so that we would have
to write
\begin{equation}
\label{eq:Jz2_App}
J_z = - p \widetilde D_T \frac{\partial T}{\partial z} - \frac{1}{\eta} \frac{\partial (k_{\mathrm{B}}T \, p)}{\partial z}
\end{equation}
with a different thermophoretic coefficient $\widetilde D_T$
as compared to Eq.~(\ref{eq:Jz_App}).
This particular form of the diffusion term results from
performing the overdamped limit in the full-fledged particle equations
of motion including the velocity degrees of freedom
(after the white noise limit has been performed first).

We can easily bring Eq.~(\ref{eq:Jz2_App}) into the form Eq.~(\ref{eq:Jz_App})
by identifying $D_T = \widetilde D_T + k_{\mathrm{B}}/\gamma$.
Based on an alternative physical reasoning,
the additional ``correction'' term $k_{\mathrm{B}}/\gamma$
has been discussed in \cite{fayolle08} to be related to the osmotic
pressure in dilute suspensions.

The corresponding difference between the Soret coefficients
$S_T=D_T/D$ and $\widetilde D_T/D$ is $1/T$. Typical values
of $S_T$ for colloidal beads are of the order of
$0.1/\mbox{K}$ to $1/\mbox{K}$
\cite{Wuerger2010,Braibanti2008,Iacopini2006}
(or even several ten or hundred per Kelvin for micron-sized particles as measured here),
so that around room temperature this difference is
expected to be negligibly small, except maybe in
temperature regions close to the sign
change of $S_T$.

We finally remark that a third variant for the
diffusion term has been discussed briefly in \cite{Parola2004},
namely $-\partial (Dp)/\partial z$. 
In principle, this prescription
amounts to yet another definition for the thermophoretic
coefficient in the probability current
\begin{equation}
\label{eq:Jz3_App}
J_z = - p \widehat D_T \frac{\partial T}{\partial z} - \frac{\partial (D \, p)}{\partial z} \, .
\end{equation}
However, as we have pointed out above, this form of the diffusion
term is not correct if $\gamma$ depends on position $z$; the difference
to the correct form $-(1/\gamma)\partial (k_{\mathrm{B}}T\,p)/\partial z$
in Eq.~(\ref{eq:Jz2_App}) is an unphysical
drift-like term $p(D/\gamma)(\partial \gamma/\partial z)$.
In case the friction coefficient $\gamma$
depends on space only
due to changes of fluid viscosity with temperature,
this term may be hidden in $\widehat D_T$,
as by comparison with Eq.~(\ref{eq:Jz2_App}) we can write
$\widetilde D_T = \widehat D_T - (D/\gamma)(\partial \gamma/\partial T)$,
or, comparing with Eq.~(\ref{eq:Jz_App}),
$D_T = \widehat D_T + k_{\mathrm{B}}/\gamma - (D/\gamma)(\partial \gamma/\partial T)$.
For water around room temperature the corresponding
``correction'' $-(1/\gamma)(\partial\gamma/\partial T)$ in the Soret
coefficient is of the order of $0.02/\mbox{K}$ and thus for colloidal particles
probably negligibly small in most cases.

Summarizing, for modeling thermophoresis in dilute particle
suspensions, one should either use Eqs.~(\ref{eq:Jz_App}) or (\ref{eq:Jz2_App})
for the probability current, while Eq.~(\ref{eq:Jz3_App}) is physically questionable.
The deviation between Eqs.~(\ref{eq:Jz_App}) and (\ref{eq:Jz2_App}) is
compensated by slightly different definitions of the
``thermal diffusion coefficient'', $D_T$ vs.\ $\widetilde D_T$.
They differ in the contribution $k_{\mathrm{B}}/\gamma$,
which is related to the temperature dependence of
the osmotic pressure in dilute suspensions \cite{fayolle08}.
Given, however, that the formulation of thermophoretic effects
is connected to the reciprocity
of heat and mass flow in temperature and density gradients
\cite{deGroot85},
the standard representation Eq.~(\ref{eq:Jz_App}) is preferable.

\balance

\footnotesize{
\bibliography{thermo} 
\bibliographystyle{rsc} 
}






\end{document}